\documentstyle[aps]{revtex}
\begin{document}
\author{Jun Xiong$^1$, De-Zhong Cao$^1,$ Feng Huang,$^1$ Hong-Guo Li,$^1$ Xu-Juan Sun%
$^1$ and Kaige Wang$^{2,1}\thanks{%
Corresponding author: wangkg@bnu.edu.cn}$}
\address{1. Department of Physics, Applied Optics Beijing Area Major Laboratory,\\
Beijing Normal University, Beijing 100875, China\\
2. CCAST (World Laboratory), P. O. Box 8730, Beijing 100080}
\title{Experimental Observation of Classical Sub-Wavelength Interference with
Thermal-Like Light }
\maketitle

\begin{abstract}
We show the experimental observation of the classical sub-wavelength
double-slit interference with a pseudo-thermal light source. The
experimental results are in agreement with the recent theoretical prediction
shown in quant-ph/0404078 (to be appeared in Phys. Rev. A).

PACS number(s): 42.50.Dv, 42.50.St, 42.25.Hz
\end{abstract}

Interference effects reflect the nature of waves, both the classical wave
and the quantum wave. For a particle, the de Broglie wavelength depends on
its mass. When two particles with the same mass combine into a whole, a
molecular, for example, the corresponding de Broglie wavelength reduces to
half of that of a single particle. Recently, a similar effect for photons,
named quantum sub-wavelength interference, has drawn attention\cite{yama}-%
\cite{shi}. For the beams generated in spontaneous parametric
down-conversion (SPDC), the interference pattern by a double-slit or a beam
splitter shows a sub-wavelength interference effect in a two-photon
observation . In comparison with massive particles, the effect was explained
by means of the photonic de Broglie wave of a multiphoton wavepacket\cite
{yama}\cite{fon1}\cite{fon2}\cite{eda}. In the further theoretical analysis,
the sub-wavelength interference effect is attributed to a quantum
entanglement of photons\cite{boto}\cite{shih2}. These studies revealed the
sub-wavelength interference as a non-classical effect occurring in the
microscopic realm. Furthermore, it has been shown theoretically that the
sub-wavelength interference effect can also exist in the macroscopic realm
in which the interfered beams generated in a high gain SPDC contain a large
number of photons\cite{na}\cite{kaige}. However, in Refs.\cite{kaige}, the
authors found another scheme to observe the double-slit sub-wavelength
interference in a joint-intensity measurement in which two photodetectors
are placed at a pair of symmetric positions with respect to the center of
double-slit. This effect occurs only in the macroscopic realm without the
microscopic counterpart. Then it has been proved in Ref. \cite{kaige1} that
the second-order spatial correlation of field is responsible for the
sub-wavelength interference and that both the entangled photon pair
generated in SPDC and the thermal light possess such spatial correlation. In
parallel, it has been found that the thermal light source can perform the
ghost imaging and interference\cite{ben1}-\cite{shih3}, which were
considered as nonclassical effects devoted by a two-photon entanglement. The
discussions were stimulated by the experimental observation of ghost imaging
and interference with classically correlated beams\cite{ben1}. Then the
authors in Ref.\cite{gigi} firstly proved that the classical correlation of
two beams obtained by splitting incoherent thermal radiation can perform the
ghost imaging. Just recently, the experimental realization of classical
correlated imaging has been reported with a pseudo-thermal light source\cite
{gigi1}\cite{shih3}.

In this paper, we report the experimental observation of the sub-wavelength
interference with a thermal-like light source. In a Young's double-slit
interference setup, we measure the intensity distribution and the
joint-intensity distribution of a thermal-like light source at the detection
plane and compare them with those of a coherent light source. The
experimental results witness that the thermal-like light can perform an
incoherent pattern in the intensity distribution and a sub-wavelength
interference pattern in the joint-intensity observation in agreement with
the theoretical predictions.

To begin with, let a coherent beam illuminate a double slit of slit width $b$
and slit distance $d$. In the interfered plane at the distance far from the
double-slit, the first-order interference-diffraction pattern (also called
one-photon double-slit interference) is described by $G^{(1)}(x,x)=A%
\widetilde{T}^2(kx/z),$ where function $\widetilde{T}$ is the Fourier
transform of the double-slit function given by 
\begin{equation}
\widetilde{T}(q)=(2b/\sqrt{2\pi })%
%TCIMACRO{\func{sinc}}
%BeginExpansion
\mathop{\rm sinc}
%EndExpansion
(qb/2)\cos (qd/2).  \label{1}
\end{equation}
$\widetilde{T}^2(kx/z)$ describes an interference-diffraction pattern with
the fringe-stripe interval $\lambda z/d,$ where $\lambda =2\pi /k$ is the
beam wavelength and $z$ is the distance between the double-slit and the
detection plane. If we perform a joint-intensity measurement in the
interfered plane, however, we obtain the second-order correlation function $%
G^{(2)}(x_1,x_2)=A^2\widetilde{T}^2(kx_1/z)\widetilde{T}^2(kx_2/z)$, which
describes a two-photon double-slit interference. In general, $G^{(2)}(x,x)$
and $G^{(2)}(x,-x)$ may exhibit two kinds of observation of two-photon
interference: the former is measured by a two-photon absorption detector and
the latter can be carried out by a joint-intensity measurement in which two
detectors are placed at a pair of symmetric positions. But for the coherent
field, the two kinds of observation show no difference because of $%
G^{(2)}(x,x)=$ $G^{(2)}(x,-x)$. Furthermore, these two kinds of two-photon
interference display the same fringe-stripe interval as that for the
one-photon interference due to the fact $G^{(2)}(x,x)=[G^{(1)}(x,x)]^2$. It
is also well known that the normalized second-order correlation function $%
g^{(2)}(x_{1,}x_2)=G^{(2)}(x_1,x_2)/[G^{(1)}(x_1,x_1)G^{(1)}(x_2,x_2)]$ is
unity which witnesses the coherence of the field.

Now, we consider one-photon and two-photon double-slit interference of a
classical thermal light source, which is assumed to radiate a monochromatic
chaotic beam $E(x,z,t)=\int E(q)\exp [iqx{\bf ]}dq{\bf \times }\exp
[i(kz-\omega t)],$ where $E(q)$ satisfies a multimode thermal statistics. In
the interfered plane, the intensity distribution and the second order
correlations are written as\cite{kaige1} 
\begin{equation}
G^{(1)}(x,x)=A\int \widetilde{T}^2(\frac{kx}z-q)S(q)dq,  \label{2}
\end{equation}
\begin{equation}
G^{(2)}(x_1,x_2)=A^2\left\{ \int \widetilde{T}^2(\frac{kx_1}z-q)S(q)dq\int 
\widetilde{T}^2(\frac{kx_2}z-q)S(q)dq+\left[ \int \widetilde{T}(\frac{kx_1}z%
-q)\widetilde{T}(\frac{kx_2}z-q)S(q)dq\right] ^2\right\} ,  \label{3}
\end{equation}
respectively, where $S(q)$ is the spatial frequency spectrum of the thermal
light. In the broadband limit, these correlations can be approximately
reduced to 
\begin{equation}
G^{(1)}(x,x)=AS(0)\widetilde{T}(0),  \label{4}
\end{equation}
\begin{equation}
G^{(2)}(x_1,x_2)=A^2S^2(0)\left\{ \widetilde{T}^2(0)+\widetilde{T}^2[\frac kz%
(x_1-x_2)]\right\} ,  \label{5}
\end{equation}
respectively. It is well known that the one-photon interference of the
thermal light disappears with a broadband spatial frequency. The random
propagation directions wash out the interference. But the interference
exists in a joint-intensity measurement even if the spatial frequency
bandwidth of the thermal fluctuation is wider. In particular, when two
detectors are placed at symmetric positions, $x$ and $-x$, the second term
of Eq. (\ref{5}), $\widetilde{T}^2(2kx/z),$ shows a sub-wavelength
interference pattern which is equivalent to the one-photon interference of a
coherent beam with half of the wavelength. In contrary to the coherent
field, the interference pattern can be also displayed in the normalized
second-order correlation. Figure 1 shows the second-order correlations $%
G^{(2)}(x,-x)$ and $g^{(2)}(x,-x)$ for both the thermal beam and the
coherent beam. The double-slit parameter $d/b=100/55$ is set so that there
are three fringe-stripes in the primary diffraction range. For a moderate
bandwidth of thermal light, the five fringe-stripes are observed as shown in
Figs. 1b and 1e. However, as for a wide bandwidth of thermal light shown in
Figs. 1a and 1d, the interference-diffraction pattern is reduced to half of
that for the coherent light. In addition, the visibility of the interference
pattern for $g^{(2)}(x,-x)$ is better than that for $G^{(2)}(x,-x)$ when the
bandwidth is not wide enough.

The experimental setup shown in Fig.2 is similar to that in Ref. \cite{shih2}
with the exception of that a pseudo-thermal light source replaces the
entangled two-photon source. The pseudo-thermal source is obtained by
injecting a focused light beam of He-Ne laser with the wavelength of 632.8 $%
nm$ into a slowly rotating (0.5Hz) ground glass disk. A double slit is
placed at a distance of 15 $mm$ from the thermal source. The two slits are
separated by $d=100$ $\mu m$ and have a width $b=55$ $\mu m$. The diffracted
radiation is split into two beams with a non-polarizing 50/50 beam splitter,
which is located at a distance of $290$ $mm$ from the double slit. The
transmitted and reflected beams are detected by small-area (0.28 $mm^2$)
Si-photodetectors D1 and D2, which are mounted on translation stages. The
distance between the detectors and the double slit is $z=550$ $mm$. The
signals from the two detectors are recorded on a digital oscilloscope
(Tektronics 3012B) for the joint-intensity measurement. By measuring the
average of the product of these two signals over a 2 seconds interval, we
obtain the average joint-intensity $\langle I_1(x_1)I_2(x_2)\rangle $, where 
$I_1(x_1)$ and $I_2(x_2)$ are light intensities detected by D1 and D2 at the
positions $x_1$ and $x_2$, respectively. As a matter of fact, the
joint-intensity correlation of the outgoing fields in the beam splitter is
proportional to a second-order correlation of the input field. For
comparison, we also measured the first-order and second-order
interference-diffraction patterns of the He-Ne laser coherent light using
the same experimental setup.

The experimental results are shown in Figs. 3-6, where Figs. 3a, 4a, 4b, 5a
and 6 show the results for the thermal light and Figs. 3b, 4c and 5b, for
the coherent light. In Fig. 3 we measure the intensity distributions of the
two outgoing beams. It is obvious that the incoherent thermal light exhibits
a diffraction pattern without fringe whereas the coherent light illustrates
a well-known interference fringe. In the theoretical simulation, we assume
that the pseudo-thermal light has a Gaussian-type spectrum $S(q)=(\sqrt{2\pi 
}w)^{-1}\exp [-q^2/(2w^2)]$. Using Eq. (\ref{2}), we can calculate the
diffraction pattern as shown by a solid line in Fig. 3a. For better fitting
to the experimental data, the normalized bandwidth of the pseudo-thermal
light is taken as $wb/(2\pi )=0.52$, which is also applied to the
theoretical simulations in Figs. 4-6. In Figs. 4 and 5, we perform the
joint-intensity measurement at positions ($x,-x$) and ($x,x$), respectively.
For the thermal light source, by measuring the normalized second-order
correlation $g^{(2)}(x,-x)$ and the second order correlation $G^{(2)}(x,-x)$%
, we can see that the sub-wavelength interference patterns are exhibited in
Figs. 4a and 4b, respectively. The fringe visibilities are 0.214 for $%
g^{(2)}(x,-x)$ and 0.160 for $G^{(2)}(x,-x)$. In Fig. 5a, the interference
disappears in the correlation $g^{(2)}(x,x)$ for the thermal light. For the
sake of comparison, $G^{(2)}(x,-x)$ and $G^{(2)}(x,x)$ of the coherent light
are measured and exhibited in Figs. 4c and 5b, respectively. However, in
Fig. 6, we measure the normalized joint-intensity correlation $g^{(2)}(x,0)$
by fixing one detector at the symmetric center and the result displays a
similar interference pattern to the one for the coherent beam without
showing the sub-wavelength effect.

The main interference features shown in Figs. 3a-5a and 6 demonstrate the
theoretical predictions in Eqs. (\ref{4}) and (\ref{5}) which are obtained
for an ideal thermal correlation with a wide bandwidth. In this case, the
maximum and minimum values of the fringe pattern should be 2 and 1,
respectively (see Figs. 1d and 1e). However, the pseudo-thermal source in
the experiment is not perfect. In addition, the photodetectors have a finite
area. Thus, we assume a modified second-order correlation $%
g^{(2)}(x_1,x_2)=1+\delta +\eta ^2\langle I_1(x_1)I_2(x_2)\rangle /[\langle
I_1(x_1)\rangle \langle I_2(x_2)\rangle ]$, where $\delta $ and $\eta $
describe the deviation from the perfect thermal correlation and detection.
The modification does not alter the main features of the interference
patterns. By taking into account the modification, the solid lines in Figs.
3a-5a and 6 indicate the theoretical simulation of the interference patterns
given by Eqs. (\ref{2}) and (\ref{3}). However, the theoretical calculations
of the interference-diffraction pattern for the coherent light are obtained
according to the function $\widetilde{T}^2(kx/z).$

The sub-wavelength interference is considered as an effect beating the
Rayleigh diffraction limit and has prospective application in
photolithography. The quantum scheme of sub-wavelength interference proposed
in Ref.\cite{shih2} is only proof-of-principle. The wavelength of the
entangled photon pair generated in the SPDC has already been doubled from
that of the pump beam. However, the intensity of an entangled photon pair is
almost negligible in any practical application. These shortcomings can be
overcome in the classical scheme. As a matter of fact, the present
experiment can be readily implemented with simple laboratory apparatus.

The experimental results are rather astonishing from a conventional
perspective of the interference. First, the interference is related to the
phase of field and the light intensity does not contain any phase
information. Second, the disorder of light source may destroy the
interference. This could be a reason why the discovery of the classical
sub-wavelength effect comes later than the quantum one. Ref.\cite{kaige1}
has interpreted the origin of both the quantum and classical sub-wavelength
interference in terms of spatial correlation of field. The correlation of
the transverse wavevectors can be formed by either the quantum entanglement
of photons or the multimode thermal statistics. In light of this, we can say
that the thermal ''disorder'' makes the intensity correlation contain the
phase information. In the early time of quantum optics when Hanbury-Brown
and Twiss claimed the temporal and spatial correlation for a stellar source%
\cite{hbt}, it was puzzling why more photon pairs are close together rather
than further part. In the photon picture, the sub-wavelength interference as
a two-photon effect is associated with photon correlation. Therefore, both
the photon correlation and the photon bunching are the nature of thermal
photon statistics.

This research was supported by the National Fundamental Research Program of
China Project No. 2001CB309310, and the National Natural Science Foundation
of China, Project No. 60278021 and No. 10074008.

Figure Captions

Fig. 1. Numerical simulations of the second-order correlation functions $%
G^{(2)}(x,-x)$ (the left side) and the normalized second-order correlation
functions $g^{(2)}(x,-x)$ (the right side) for both the thermal light (a, b,
d, and e) and the coherent light (c and f). The ratio between slit distance
and slit width is $d/b=100/55$. The spatial frequency spectrum $S(q)$ of the
thermal light is assumed to be a Gaussian function with a bandwidth $w$. In
(a) and (d) a wide bandwidth $wb/(2\pi )=10$ is taken, and in (b) and (e) a
moderate bandwidth $wb/(2\pi )=0.52$ is taken for the thermal light.

Fig. 2. Sketch of the experimental setup.

Fig. 3. Average intensity distributions of the two outgoing beams from the
beam splitter for (a) the pseudo-thermal light and (b) the coherent light.
The experimental data are indicated by triangles and circles detected by D1
and D2, respectively. In Figs. 3-6, the solid lines show the numerical
simulations in accordance with the experimental setup, and the bandwidth of
the thermal light spectrum is taken as $wb/(2\pi )=0.52$.

Fig. 4. Interference-diffraction patterns by measuring (a) the normalized
second-order correlation function $g^{(2)}(x,-x)$ and (b) the second-order
correlation function $G^{(2)}(x,-x)$ of the pseudo-thermal light, whereas in
(c) $G^{(2)}(x,-x)$ is measured for the coherent light. In Figs. 4-6, the
experimental data are indicated by square-dots and the modification
parameters $\delta =0.04$ and $\eta =0.66$ are taken in the numerical
simulations for the thermal light.

Fig. 5. Interference-diffraction patterns by measuring (a) the normalized
second-order correlation function $g^{(2)}(x,x)$ of the pseudo-thermal light
and (b) the second-order correlation function $G^{(2)}(x,x)$ of the coherent
light.

Fig. 6. Interference-diffraction patterns by measuring the normalized
second-order correlation function $g^{(2)}(x,0)$ of the pseudo-thermal light.

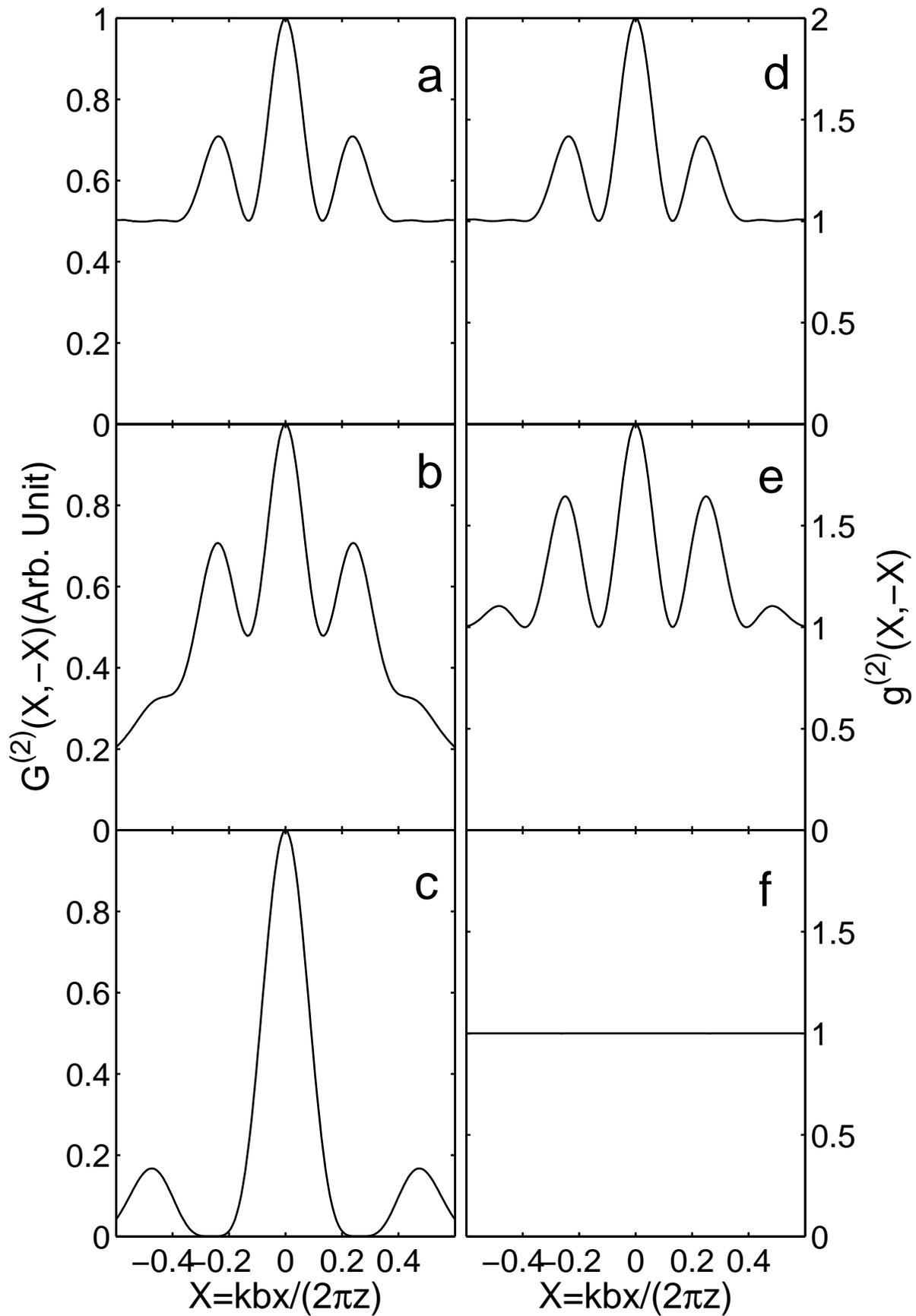
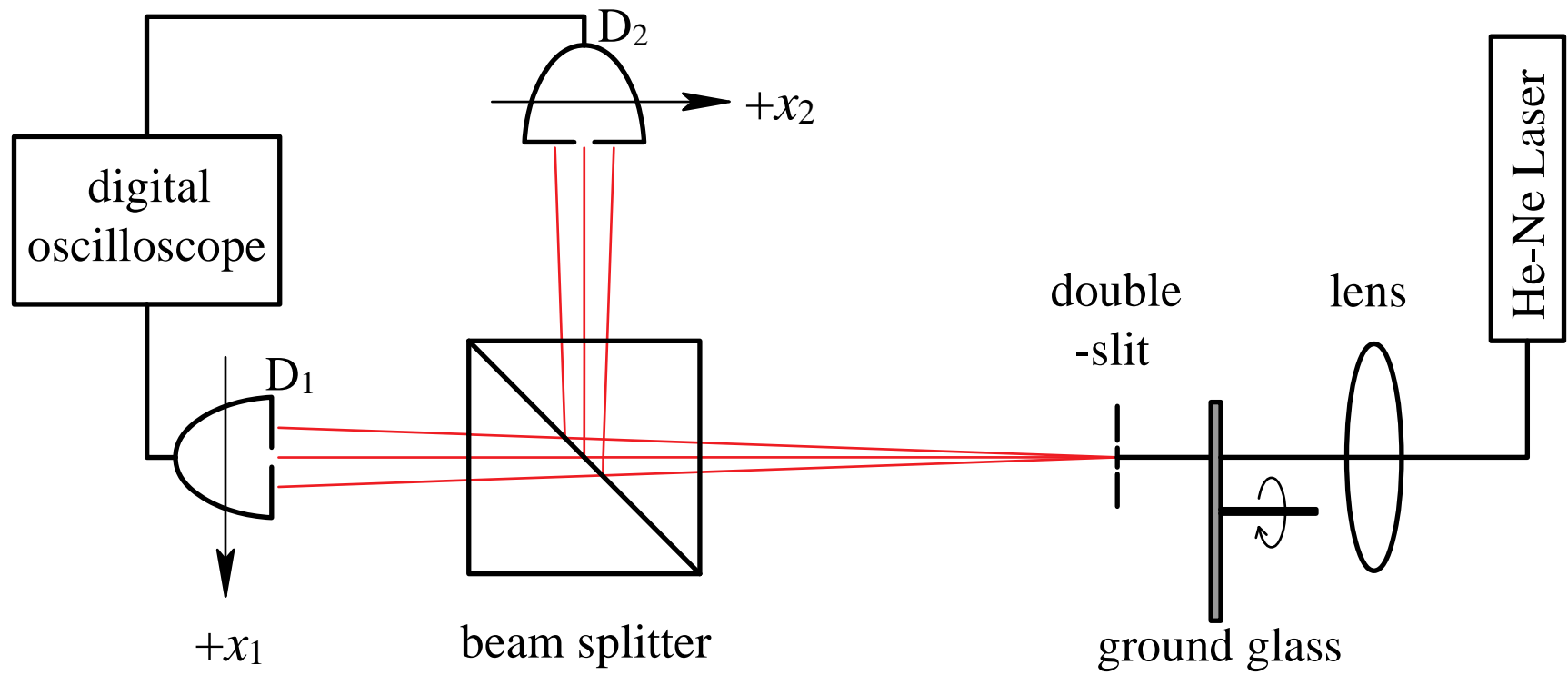
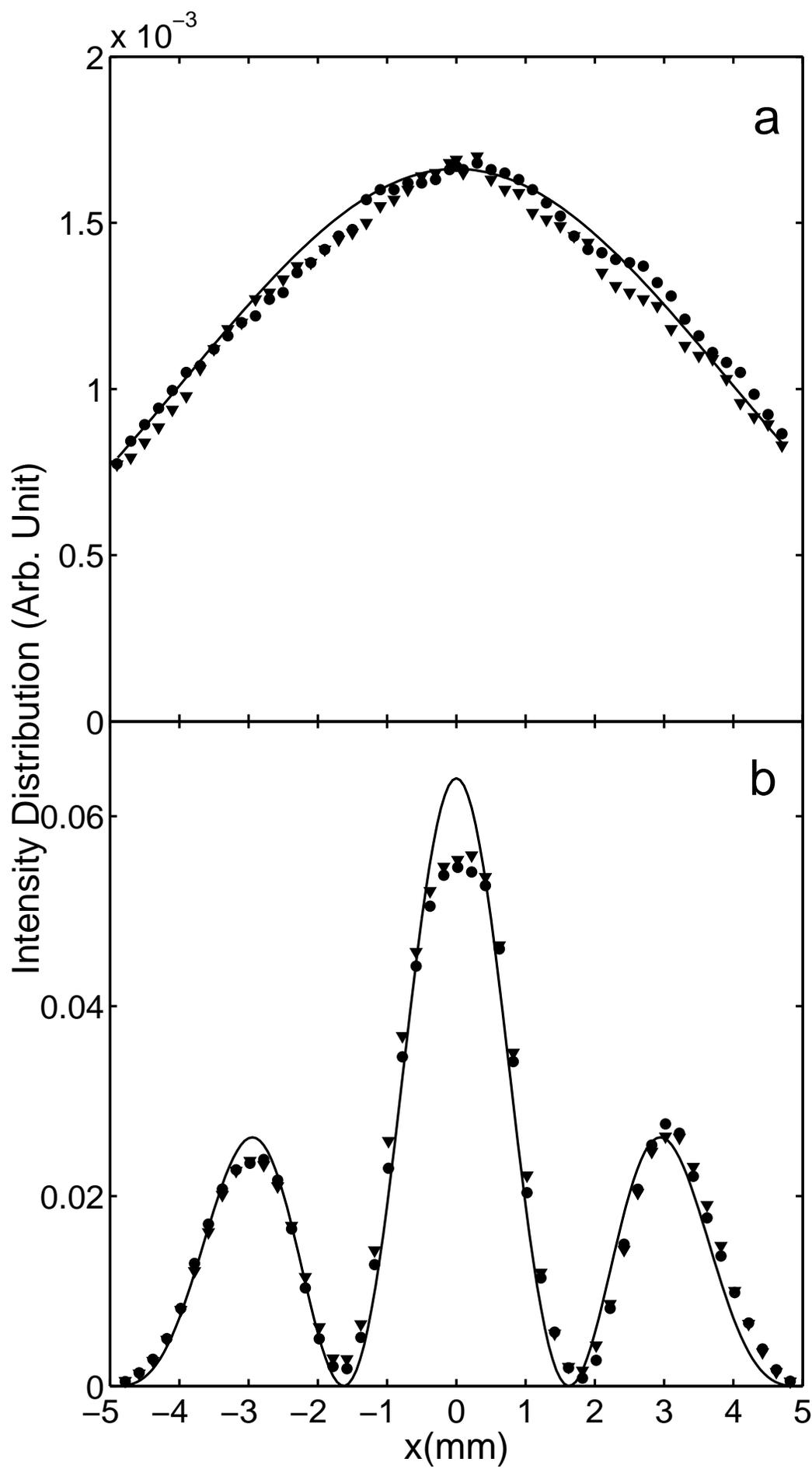
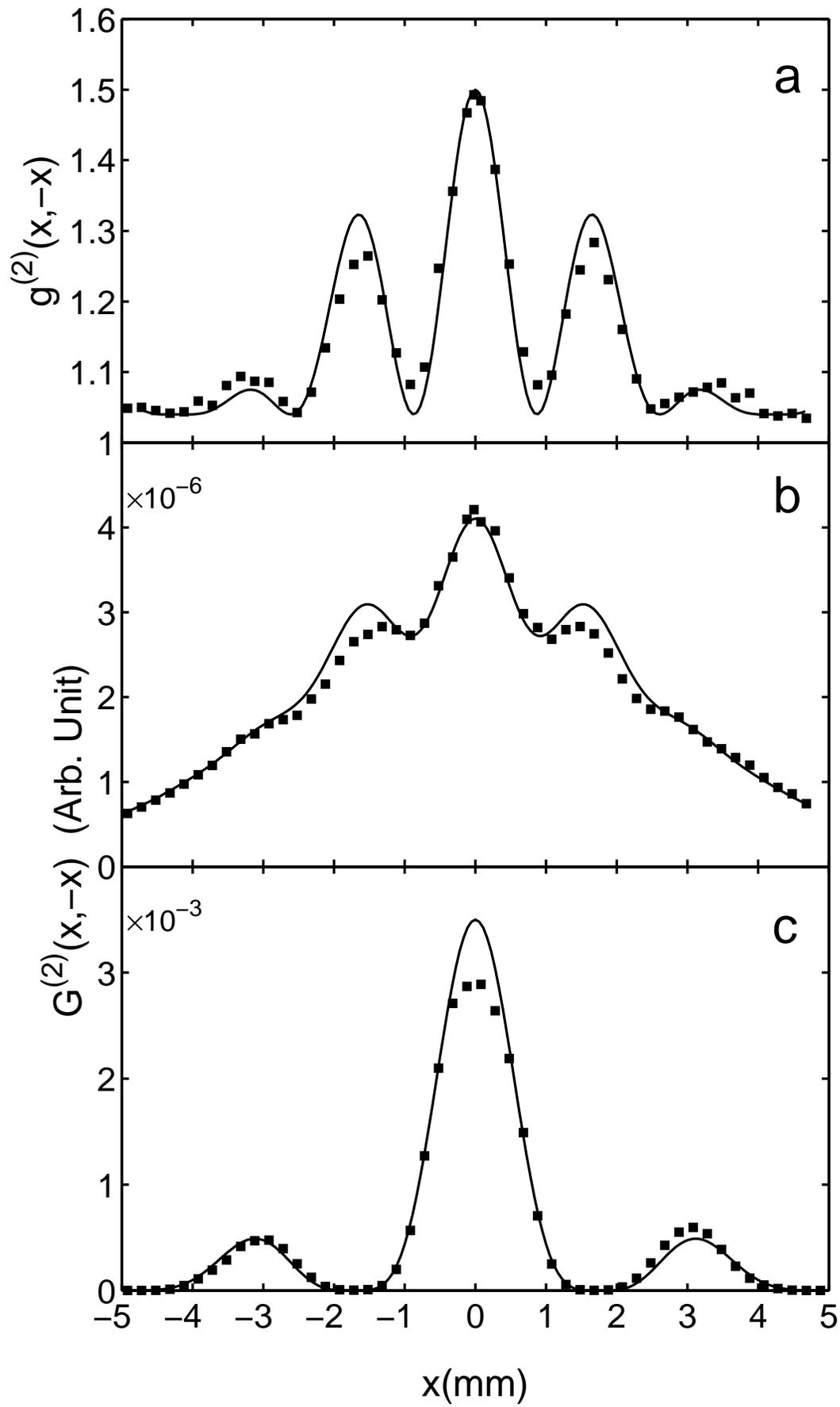
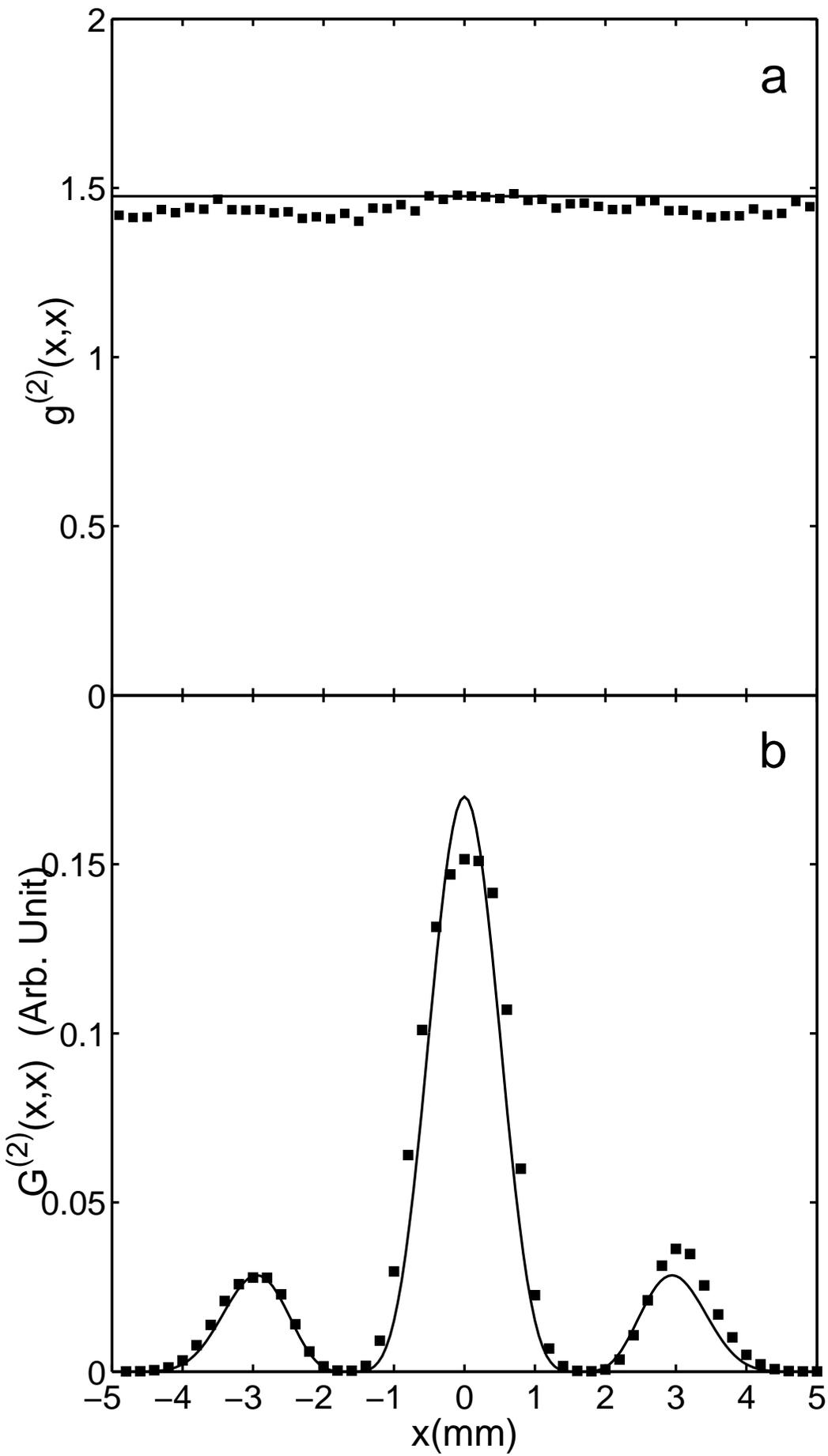
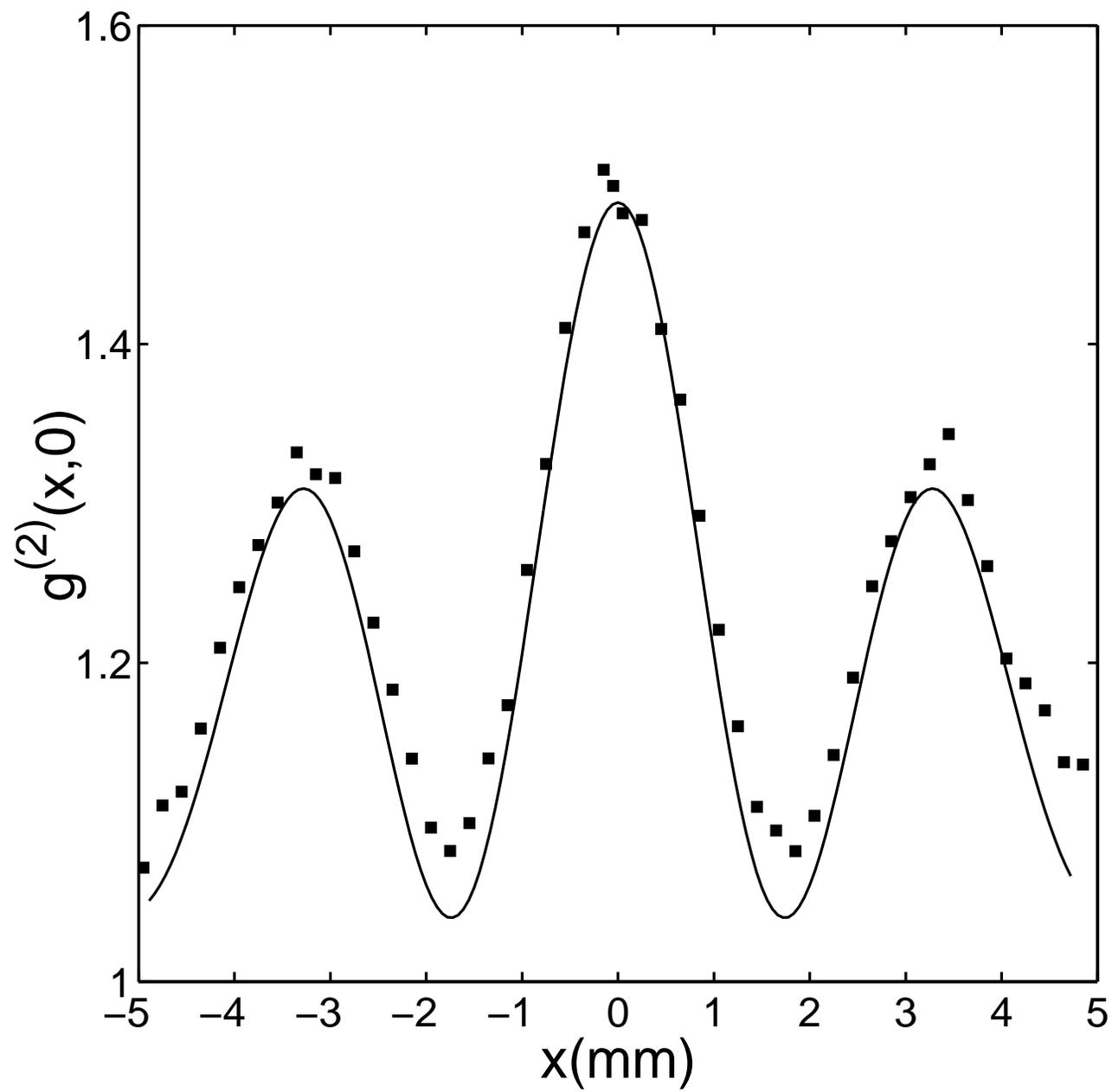
\end{document}